\documentclass[12pt,fleqn]{iopart}

\usepackage{iopams}

\usepackage[numbers]{natbib}
\usepackage{xparse}
\NewDocumentCommand{\multicite}{>{\SplitList{,}}m}{%
  \ProcessList{#1}{\cite}%
}

\usepackage{soul}
\usepackage{xcolor}

\usepackage{graphicx}

\begin{document}

\title[Neural Importance Resampling]{Neural Importance Resampling: A Practical Sampling Strategy for Neural Quantum States}

\author{Eimantas Ledinauskas\textsuperscript{1,2} \& Egidijus Anisimovas\textsuperscript{1}}

\address{{\bf 1} Institute of Theoretical Physics and Astronomy, Vilnius University\\
Saul\.{e}tekio al.\ 3, LT-10257, Vilnius, Lithuania \\ {\bf 2} Baltic Institute of Advanced Technology, Pilies St. 16-8, LT-01403, Vilnius, Lithuania}
\ead{eimantas.ledinauskas@ff.vu.lt, egidijus.anisimovas@ff.vu.lt}
\vspace{10pt}
\begin{indented}
\item[] \today
\end{indented}

\begin{abstract}
Neural quantum states (NQS) have emerged as powerful tools for simulating many-body quantum systems, but their practical use is often hindered by limitations of current sampling techniques. Markov chain Monte Carlo (MCMC) methods suffer from slow mixing and require manual tuning, while autoregressive NQS impose restrictive architectural constraints that complicate the enforcement of symmetries and the construction of determinant-based multi-state wave functions. In this work, we introduce Neural Importance Resampling (NIR), a new sampling algorithm that combines importance resampling with a separately trained autoregressive proposal network. This approach enables efficient and unbiased sampling without constraining the NQS architecture. We demonstrate that NIR supports stable and scalable training, including for multi-state NQS, and mitigates issues faced by MCMC and autoregressive approaches. Numerical experiments on the 2D transverse-field Ising and $J_1$-$J_2$ Heisenberg models show that NIR outperforms MCMC in challenging regimes and yields results competitive with state of the art methods. Our results establish NIR as a robust alternative for sampling in variational NQS algorithms.
\end{abstract}

%
%
\submitto{Machine Learning: Science and Technology}
%
%
%

\section{Introduction}

Neural quantum states (NQS), originally introduced by \cite{Carleo_first_NQS}, have seen rapid methodological advancements, enabling their successful application to an increasingly diverse range of many-body quantum systems \multicite{carra_review_2020, nqs_review_medvidovic_2024, nqs_review_lange_2024, nqs_systematic_review_2022}.  Today, NQS are considered state-of-the-art tools for tackling the complexities of many-body quantum physics.

Methods that utilize NQS typically require sampling configurations according to the probability distribution defined by the NQS wave function. Given the exponentially large dimensionality of many-body quantum systems, this sampling is commonly performed using Markov chain Monte Carlo (MCMC) techniques, particularly the Metropolis-Hastings (MH) algorithm. However, this approach has several well-known drawbacks, including slow mixing, autocorrelation, difficulties in equilibration, and susceptibility to local traps. Additionally, effective sampling often relies on careful tuning of the proposal distribution.

These issues can be addressed using autoregressive NQS, which enable exact and efficient sampling of basis states \multicite{rnn_nqs_allah, autoregressive_CNN_NQS}. However, this approach imposes architectural constraints on the neural networks, introducing new limitations. For instance, enforcing global constraints and symmetries of the wave function becomes more challenging and requires special approaches which can be suboptimal \cite{nqs_design_reh2023}. Another limitation of autoregressive NQS arises in the context of a recent method proposed by \cite{determinant_nqs_2024}, which aims to approximate multiple lowest energy states. This method involves variational Monte Carlo with a multi-state wave function constructed as a determinant of multiple single-state NQS. It remains unclear how to perform exact sampling from such a determinant wave function, even when each constituent NQS is autoregressive.

In this work, we introduce an alternative sampling algorithm based on importance resampling, with proposals generated by a separate autoregressive neural network. This approach employs a dedicated sampling neural network (SNN) that automatically learns the probability distribution defined by the NQS, eliminating the need for manually designed proposal distributions. The learned proposal distribution is efficient enough for importance resampling to be viable in the high-dimensional spaces typical of many-body quantum systems. We argue that this algorithm is practical for NQS applications and offers advantages over both MCMC and autoregressive NQS sampling. Compared to MCMC, importance resampling is simpler, makes fewer assumptions, and is therefore more robust. Additionally, thanks to the trainable SNN, it requires less manual tuning for specific problems. In contrast to autoregressive NQS, our method does not impose architectural constraints on the NQS itself, since sampling is delegated to a separate network. This allows for straightforward enforcement of constraints and facilitates the use of multi-state determinant NQS.

The remainder of this paper is organized as follows. In Sec. \ref{sec:context}, we provide a brief overview of the necessary background and context. Sec. \ref{sec:NIR} introduces the proposed sampling method in detail and discusses the related work. In Sec. \ref{sec:numerical_experiments}, we present numerical experiments that demonstrate the practicality and effectiveness of our approach. Finally, Sec. \ref{sec:conclusions} offers concluding remarks.

\section{Context}
\label{sec:context}

\subsection{Single-state NQS}

To streamline presentation, we focus in this work on systems with discrete degrees of freedom, such as spin lattices or tight-binding models. These systems can be described by a wave function, which is a complex vector of the form:
\begin{equation}
  |\psi\rangle = \sum_s \psi(s) |s\rangle.
\end{equation}
Here, $|s\rangle$ are the basis vectors in the computational basis defined as the tensor product of the states at a single site and $\psi(s) = \langle s | \psi \rangle$ are the corresponding components of the state vector. In many-body systems, the dimension of this vector grows exponentially with system size, quickly becoming intractable to handle directly.

NQS use a neural network with parameters $\theta$ to represent the wave function by mapping $|s\rangle$ to the corresponding components $\psi_\theta(s)$. This approach compresses the otherwise intractable wave function vector into a more manageable neural network representation. The parameters $\theta$ can be optimized for specific tasks. NQS have been widely applied in various contexts, including: ground state approximation via variational Monte Carlo (VMC), quantum state tomography, and time evolution \multicite{carra_review_2020, nqs_review_medvidovic_2024, nqs_review_lange_2024, nqs_systematic_review_2022}.

\subsection{Multi-state NQS}
\label{sec:multistate_nqs}

\cite{determinant_nqs_2024} proposed a generalization of VMC that enables the simultaneous search for multiple low-energy eigenstates. This is achieved by transforming the problem of finding excited states of a given system into the problem of finding the ground state of an expanded system. While the method can be used with various types of variational ansatzes, it is particularly well-suited to NQS. The approach requires the variational wave function to take the following form:
\begin{equation}
    \mathbf{\Psi}(s^1, \ldots, s^K) = \det\left(
    \begin{array}{ccc}
        \psi_1(s^1) & \cdots & \psi_K(s^1) \\
        \vdots & \ddots & \vdots \\
        \psi_1(s^K) & \cdots & \psi_K(s^K)
    \end{array} \right)
\end{equation}
Here, $s^1, \ldots, s^K$ is a tuple of $K$ basis states of the original system, and $\psi_j(s^k)$ are single state ansatzes (neural networks in our case). The energy is generalized from a scalar to a matrix as follows:
\begin{equation}
    \label{eq:energy_matrix}
    \mathbf{E}_{\Psi} = 
    \left\langle\left(\begin{array}{ccc}
        \psi_1(s^1) & \cdots & \psi_K(s^1) \\
        \vdots & \ddots & \vdots \\
        \psi_1(s^K) & \cdots & \psi_K(s^K)
    \end{array}\right)^{-1}
    \left(\begin{array}{ccc}
        \hat{H}\psi_1(s^1) & \cdots & \hat{H}\psi_K(s^1) \\
        \vdots & \ddots & \vdots \\
        \hat{H}\psi_1(s^K) & \cdots & \hat{H}\psi_K(s^K)
    \end{array}\right)\right\rangle _{s^1, \ldots, s^K \sim \Psi^2}
\end{equation}
Here, $\langle \ldots \rangle_{s^1, \ldots, s^K \sim \Psi^2}$ denotes the expectation value over tuples of basis states sampled according to the probability distribution defined by $\mathbf{\Psi}^2$ and $\hat{H}$ is the Hamiltonian operator of the original system. The trace of the energy matrix, which is minimized during VMC, represents the total energy of the $K$ states. Estimates of the individual lowest-energy eigenvalues can be obtained by diagonalizing the matrix after VMC convergence.

\subsection{Observable and gradient estimation}

Because the full wave function vector is intractable, the expectation values of observables must be estimated stochastically. This is accomplished using the following well-known identities:
\begin{equation}
    O_{\mathrm{loc}}(s) = \sum_{s'} \frac{\psi(s')}{\psi(s)} O_{s,s'}
\end{equation}
\begin{equation}
    \left\langle \hat{O} \right\rangle = \left\langle O_{\mathrm{loc}}(s) \right\rangle \approx \frac{1}{N} \sum_{s \sim \psi^2} O_{\mathrm{loc}}(s)
\end{equation}
Here, $O_{\mathrm{loc}}(s)$ denotes the local estimator of the operator $\hat{O}$, and $O_{s,s'} = \langle s|\hat{O} |s' \rangle$ are the matrix elements of the operator. This method is feasible only when $O_{s,s'}$ is sparse, allowing efficient computation of $O_{\mathrm{loc}}(s)$. The approximation relies on sampling basis states $s$ according to the probability distribution defined by the squared amplitude of the NQS wave function.

One particularly important observable is the energy, which is minimized during VMC to obtain the ground state. The gradient of the energy with respect to the NQS parameters---necessary for NQS optimization---can be computed using the following identity:
\begin{equation}
    \frac{\partial E}{\partial\theta_j} = 2\Re\left\{ \left\langle H_{\mathrm{loc}}(s)  \frac{\partial}{\partial\theta_j}\log\psi^{*}(s) \right\rangle -\left\langle H_{loc}(s)\right\rangle \left\langle \frac{\partial}{\partial\theta_j}\log\psi^{*}(s)\right\rangle \right\} \,.
\end{equation}
As with the expectation values of the operators, evaluating this gradient requires sampling from the NQS distribution.

\subsection{MCMC sampling}

One commonly used method to sample basis states from NQS is the Metropolis-Hastings (MH) algorithm \multicite{Metropolis1953, Hastings1970}. MH constructs a Markov chain with transition probabilities defined by a proposal distribution $T(s \rightarrow s')$ and an acceptance ratio:
\begin{equation}
    A(s \rightarrow s') = \min\left(1, \frac{|\psi(s')|^2 T(s' \rightarrow s)}{|\psi(s)|^2 T(s \rightarrow s')}\right)
\end{equation}
Under the conditions of irreducibility, aperiodicity, and detailed balance, this procedure guarantees convergence to the desired stationary distribution.

The choice of proposal distribution $T$ strongly influences sampling efficiency. While simple local proposals are often sufficient, some specific systems require nontrivial or learned proposals. Dynamic proposals, such as those parameterized by neural networks, can adapt to the structure of the target distribution and improve mixing. However, ensuring proper MCMC convergence with dynamic proposal distributions imposes some nontrivial constraints. In particular, dynamically learned or adaptive proposals must be designed to preserve the ergodicity and detailed balance (or at least stationarity) of the Markov chain. One essential constraint is diminishing adaptation, which requires that the magnitude of changes to the proposal mechanism decrease over time, ensuring that the chain stabilizes and avoids persistent non-stationarity \cite{dynamic_mcmc_proofs_Roberts2007}. Another is bounded convergence, which requires that the adapted proposals remain within a controlled family that maintains uniform ergodicity \cite{dynamic_mcmc_proofs_Andrieu2007}.

\subsection{Autoregressive NQS}
\label{sec:autoregressive_nqs}

Another common approach to sampling from NQS is to use autoregressive neural networks \multicite{rnn_nqs_allah, autoregressive_CNN_NQS}. Any probability distribution over a discrete set of variables can be factorized into a product of conditional probabilities. Analogously, wave functions $\psi(s)$ defined on discrete configurations $s = (s_1, \ldots, s_N)$---where each $s_j$ may represent, for example, a quantum number associated with a lattice site---can be written in the form:
\begin{equation}
    \psi(s) = \prod_{j=1}^N \psi_j(s_j \,|\, s_1, \ldots, s_{j-1})
\end{equation}
These conditional amplitudes can be modeled using an autoregressive neural network, which enables exact sampling from $|\psi(s)|^2$ by sequentially generating each $s_j$ from the corresponding conditional distribution. This eliminates the need for MCMC.

Autoregressive NQS offer efficient and unbiased sampling but have certain limitations. The autoregressive structure makes it more difficult to enforce global constraints and symmetries. For example, \cite{nqs_design_reh2023} found that feed-forward NQS outperform autoregressive NQS in modeling the $J_1$-$J_2$ model, primarily due to a suboptimal symmetrization procedure imposed by the autoregressive structure. Moreover, it is unclear how to extend the autoregressive framework to support the multi-state NQS construction described in Sec.~\ref{sec:multistate_nqs}. Even when single-state wave functions are modeled autoregressively, it remains unresolved how this structure can be leveraged to sample from the full determinant-based ansatz.

\section{Neural importance resampling}
\label{sec:NIR}

\subsection{Importance resampling}

Importance resampling (IR), also known as sampling-importance resampling \cite{importance_resampling_Rubin_1987}, is an algorithm used to obtain samples from a target distribution $p(s)$ by leveraging samples from a proposal distribution $q(s)$. First, a batch of $N_q$ samples $s^j$ is drawn from $q(s)$. Then, importance weights are computed for each sample as $w_j = p(s^j) / q(s^j)$. Finally, a new batch of $N_p$ samples is drawn from the original batch, with selection probabilities proportional to the importance weights. This method is simple to implement and easy to understand. However, its main limitation is that, in high-dimensional settings, sampling efficiency requires the proposal distribution to overlap significantly with the target distribution. Designing such a proposal distribution is often challenging, which usually restricts the method’s applicability to low-dimensional problems.

We note that if sampling is used solely to compute expectations under the target distribution, then importance sampling can be applied by simply reweighting samples drawn from the proposal distribution when computing averages. This approach should perform similarly to resampling. In this work, we chose to use resampling purely for implementation reasons: It allows us to replace only the sampler component of an existing computation pipeline without modifying the rest of the code.

\subsection{Autoregressive proposal network}
\label{sec:proposal_network}

As mentioned in the previous section, importance resampling (IR) requires substantial overlap between the proposal and target distributions. We propose that, in high-dimensional settings, this overlap can be achieved by using an autoregressive neural network as the proposal distribution. To ensure that the proposal remains close to the target, the network is trained prior to sampling. In this work, we use a Transformer architecture \cite{transformers_Vaswani_2017} (with reordered layer normalization \cite{transformer_layernorm_Xiong_2020}), motivated by the success of Transformers as multi-domain generative models \cite{LLMs_review_Raiaan2024}, which demonstrates their ability to learn highly complex distributions. The approach for sampling in this case is identical to the autoregressive NQS described in Sec. \ref{sec:autoregressive_nqs}, except that the neural network now outputs the probability directly instead of the complex amplitude. Each site configuration is passed as a separate input token, encoded as a learnable embedding vector of dimension $d_{embd}$. Learnable positional embeddings are added to these token embeddings. The resulting vectors are then processed by standard transformer layers using a causal attention mask. Each output vector is subsequently linearly projected to a scalar value, which represents the conditional probability logits. After applying a softmax, these values are interpreted as $p(s_j|s_1,...,s_{j-1})$. To ensure that the prediction for $s_j$ depends only on the previous site configurations $s_1,...,s_{j-1}$, the input token sequence is shifted. This is achieved by prepending a learnable token to the beginning of the input sequence and discarding the final token.

\subsection{Forward Kullback–Leibler divergence loss}

Maximizing the overlap between the proposal and the target distributions can be achieved by minimizing the Kullback–Leibler divergence (KLD). There are two variants of KLD: the forward (mean-seeking) form,
\begin{equation}
    L_\mathrm{f} = \sum_s p(s) \log\frac{p(s)}{q(s)} \,,
\end{equation}
and the backward (mode-seeking) form,
\begin{equation}
    L_\mathrm{b} = \sum_s q(s) \log\frac{q(s)}{p(s)} \,.
\end{equation}
The gradients of these objectives can be estimated using the following expressions:
\begin{equation}
    \frac{\partial L_\mathrm{f}}{\partial\theta_j} = -\left\langle \frac{\partial}{\partial\theta_j}\log q(s)\right\rangle _{s\sim p}\,,
\end{equation}
\begin{equation}
   \frac{\partial L_\mathrm{b}}{\partial\theta_j} = \left\langle \left(\log q(s)-\log p(s)\right) \frac{\partial\log q(s)}{\partial\theta_j} \right\rangle _{s\sim q}\,.
\end{equation}
The gradient of the backward KLD is easier to estimate, as it only requires sampling from $q(s)$, which can be done efficiently and without bias using the autoregressive sampling network. However, the forward KLD is better aligned with the requirement of importance resampling (IR) that $q(s) > 0$ whenever $p(s) > 0$ due to its mean-seeking property. The downside of forward KLD is that it requires samples drawn from $p(s)$ (the same resampled configurations that are used for NQS optimization), which can be biased if the overlap between the NQS and sampling network distributions is not yet sufficient. In our numerical experiments, we found that both loss functions are effective, but forward KLD results in more robust training of NQS, with reduced sensitivity to hyperparameters. Training with the backward KLD often results in mode collapse, where certain subsets of configurations with high $p(s)$ have very low $q(s)$ and are rarely proposed, preventing the IR step from correcting this bias. This behavior is visible in the effective sample size (see Sec. \ref{sec:adaptive_retraining}, which occasionally spikes when a configuration with very small $q(s)$ is nevertheless sampled, producing very large importance weights. We observed these spikes when using the backward KLD and switched to the forward KLD, after which the issue disappeared. To further enforce the requirement that $q(s) > 0$ whenever $p(s) > 0$, we add a small floor $p_\mathrm{floor}$ to the single-site probabilities, ensuring that all configurations satisfy $q(s) > 0$.

In this work, we focus on the multi-state NQS described in Sec.~\ref{sec:multistate_nqs}, which is permutation-invariant with respect to the order of the input configurations $ s^1, \ldots, s^K $ by construction. To encourage the proposal network to learn this property, we randomly permute the order of the states before using them for training. This approach could be extended to incorporate other types of symmetries, depending on the Hamiltonian under study.

\subsection{Adaptive proposal network retraining}
\label{sec:adaptive_retraining}

The effectiveness of NIR critically depends on the overlap between the target distribution, given by the NQS, and the proposal distribution produced by the sampling network. However, during training, this overlap can degrade rapidly if the NQS and the sampling network begin to diverge. In such cases, the NIR resampling procedure can fail to adequately correct for the mismatch between distributions. To mitigate this issue, we monitor the effective sample size $ESS = \left(\sum_j w_j \right)^2 / \sum_k w_k^2$. We sample configurations from the proposal network until $ESS$ exceeds a predefined threshold $\alpha_{ESS}$. If the overlap with the target distribution is poor, more proposals must be generated. This compensates for low overlap and results in stable training with high-quality sample batches. After that, importance resampling is used to obtain samples from the NQS. These samples are then used to train the proposal network with the forward KLD loss. 

\begin{figure}
    \centering
    \includegraphics[width=\textwidth]{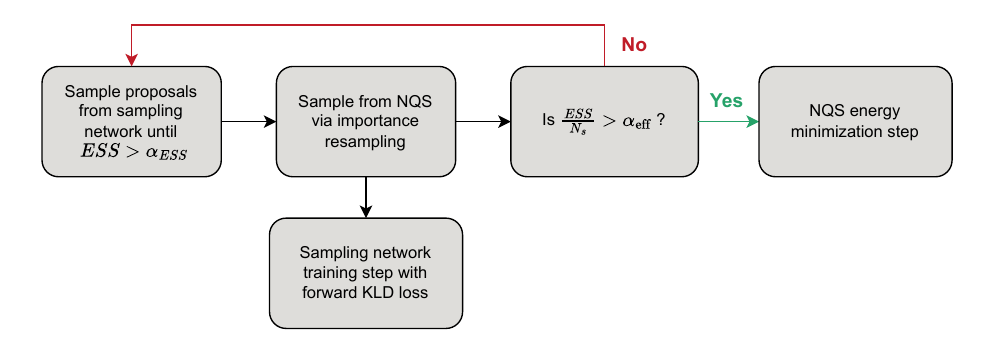}
    \caption{ Scheme for the proposed NQS training algorithm. \label{fig:nir_scheme}}
\end{figure}

However, if the overlap between the proposal network and NQS distributions is low, sampling can become very slow. To prevent divergence between the two distributions, we introduce a second safeguard: the obtained samples are used to train the NQS only if the sampling efficiency, defined as the ratio between ESS and the total number of proposed configurations, exceeds a predefined threshold $\alpha_\mathrm{eff}$. If the efficiency is too low, we repeat the sampling and proposal-network-training steps. This provides a simple mechanism for adaptive retraining and helps maintain alignment between the distributions during optimization. A block diagram for this algorithm is illustrated in Fig. \ref{fig:nir_scheme}.

\subsection{Related work}

Recent work across machine learning and computational science communities has shown that importance resampling becomes practical in high‑dimensional spaces once the proposal distribution is learned by a neural network. Neural Adaptive Sequential Monte Carlo introduced gradient‑based adaptation of recurrent proposal networks, minimizing the forward-KLD to the target and repeatedly resampling particles to maintain diversity \cite{Gu_NASMC_2015}. Building on this idea, Auto‑Encoding SMC integrates proposal learning with variational inference, treating resampling weights as part of an evidence lower bound that is optimized jointly with generative and proposal networks \cite{Le_AESMC_2018}. In computer graphics, Neural Importance Sampling employs normalizing flows to learn globally coherent proposals that drastically reduce Monte Carlo variance when integrating high‑dimensional light‑transport integrands \cite{Mueller_NIS_2019}. Flow-based proposals have also been used for event generators in particle physics, where exhaustive neural importance sampling yields unbiased estimates with competitive efficiency \cite{PinaOtey_ENIS_2020}.

At the same time, generative neural samplers have been applied in statistical physics. For instance, \cite{neural_samplers_for_classical_Ising} propose a framework for asymptotically unbiased estimation of observables (including partition-function-dependent ones) using generative neural samplers. Autoregressive neural networks have also been used as adaptive proposals in MCMC for spin-glass models, yielding improved efficiency \cite{neural_proposal_mcmc_spin_glasses}. More broadly, adaptive MCMC methods augmented with normalizing-flow transitions have been shown to learn nonlocal proposal kernels and accelerate sampling in high-dimensional spaces \cite{normalizing_flow_mcmc_continuous}. Recent work further explores this interface between flow-based proposals and energy-based modeling: adaptive flow samplers have been coupled with energy-based models to maintain balanced learning dynamics between data-driven and model-driven components \cite{grenioux2023balanced}, while systematic benchmarks in statistical-physics settings have clarified when machine-learning-assisted Monte Carlo methods truly outperform conventional algorithms \cite{ml_assisted_mc_statphys_2025}.

Collectively, these studies demonstrate that neural proposal models trained with KL‑based objectives can achieve substantial overlap with the target distribution required by importance resampling, motivating our adoption of a dedicated sampling neural network in the NIR algorithm.

\section{Numerical experiments}
\label{sec:numerical_experiments}

\subsection{Transverse field Ising model}

In this section, we present the results of numerical experiments performed on the paradigmatic two-dimensional transverse-field Ising (TFI) model. The Hamiltonian for this system is given by:
\begin{equation}
    \hat{H} = -\sum_{\langle j,k\rangle} \hat{\sigma}_z^j \hat{\sigma}_z^k - g \sum_j \hat{\sigma}_x^j,
\end{equation}
where $\langle j,k\rangle$ denotes pairs of nearest-neighbor sites on the lattice, $\hat{\sigma}_x^j$ and $\hat{\sigma}_z^j$ are Pauli matrices acting on site $j$, and $g$ is the strength of the transverse field. The first term favors spin alignment along the $z$-axis, promoting the ferromagnetic order, while the second term controls quantum fluctuations that compete with this order and drive the system towards the paramagnetic phase. On an infinite square lattice the phase transitions between these regimes occurs at $g\approx3.0444$ \cite{TFI_phase_transition_2002}, and due to the quantum-to-classical mapping the transition belongs to the 3D classical-Ising universality class. Thus, the TFI model provides a very suitable benchmark problem that combines conceptual simplicity with nontrivial behavior.

The largest TFI lattice considered in this work is $8\times8$, which may seem small for modern methods. However, to target multiple low-energy states the Hilbert space must be extended (see Sec. \ref{sec:multistate_nqs}). For three states, the system effectively has 192 sites, which is comparable in size to a $14\times14$ lattice.

\subsection{Multi-state NQS details}

We search for the three lowest energy states, which requires three separate neural networks to construct the multi-state NQS described in Sec. \ref{sec:multistate_nqs}. We use a multilayer perceptron (MLP) architecture for these networks. Each network consists of $n_{\mathrm{layer}} = 4$ hidden layers, with each layer containing $d_{\mathrm{mlp}} = 256$ neurons. Every linear projection is followed by layer normalization \cite{layer_norm} and the GELU activation function \cite{gelu_activation}. The architecture of the sampling proposal network is described in Sec. \ref{sec:proposal_network}. We use ADAM optimizer \cite{adam_paper} for both the NQS and the sampling network. For the NQS, we also employ the minSR method \cite{minSR_Chen_2024} method to improve convergence. A table of hyperparameters can be found in Sec. \ref{sec:appendix_hyperparams}. The NQS code was implemented using the JAX \cite{jax_github}, Equinox \cite{equinox_2021}, and Optax \cite{deepmind_jax_ecosystem} libraries.

\subsection{Comparing NIR with MCMC}

As previously noted in \cite{autoregressive_CNN_NQS} and \cite{TFI_phase_transition_2022}, the TFI poses challenges for NQS with an MCMC sampler in the ferromagnetic phase. The primary difficulty lies in the probability distribution of the degenerate ground state, which is sharply peaked, with the two most probable configurations (corresponding to all spins up and all spins down) being far apart in configuration space.

\begin{figure}
    \centering
    \includegraphics[width=\textwidth]{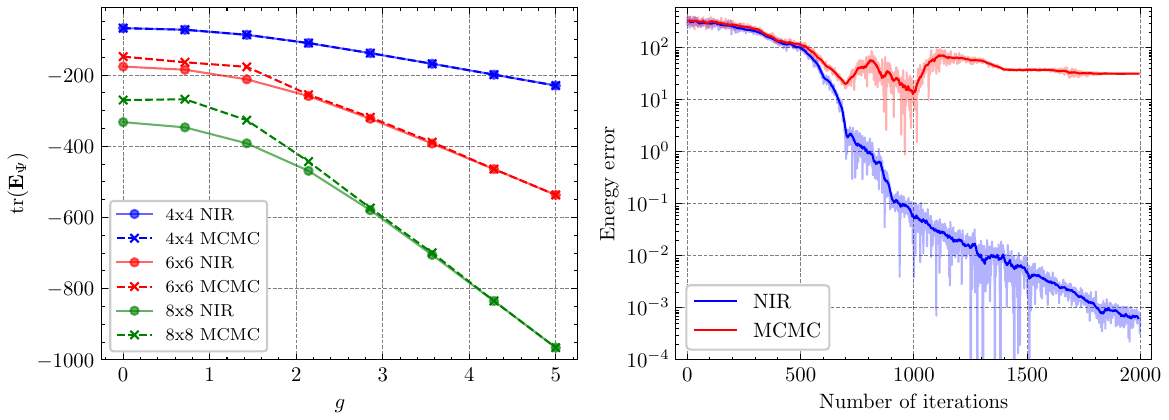}
    \caption{\textbf{Left:} Dependence of the trace of the final variational energy matrix on transverse field strength $g$ for three lattice sizes ($4\times4$, $6\times6$, $8\times8$). For every combination, two data points are shown, differing only in the sampler used. \textbf{Right:} Example of the error evolution in the sum of the three lowest energies during a training run of NQS using NIR (blue) and MCMC (red) for TFI on an $8 \times 8$ lattice and $g = 0.1$. The transparent curves show the values obtained at each iteration from the batch used in the optimization step, and the solid lines represent their moving averages.} \label{fig:mcmc_vs_nir}
\end{figure}

We performed multiple pairs of identical runs: one using the NIR sampler and the other using the MCMC sampler, keeping all other hyperparameters fixed. The comparison of the final variational energy (the trace of the energy matrix defined in Eq. (\ref{eq:energy_matrix})) across different transverse-field strengths and lattice sizes is shown in the left panel of Fig. \ref{fig:mcmc_vs_nir}. The data points correspond to averages over three runs with different random seeds. On small lattices, such as $4\times4$, both methods perform similarly. However, in the ferromagnetic regime, a performance gap appears already on the $6\times6$ lattice and becomes more pronounced on the $8\times8$ lattice. With the MCMC sampler, training often becomes unstable, and the NQS frequently gets trapped in an excited state with energy significantly above the lowest three. In contrast, NIR maintains stable training. The right panel of Fig. \ref{fig:mcmc_vs_nir} shows the typical evolution of the energy error during training for both methods with $8\times8$ lattice and $g=0.1$.

\begin{figure}
    \centering
    \includegraphics[width=0.5\textwidth]{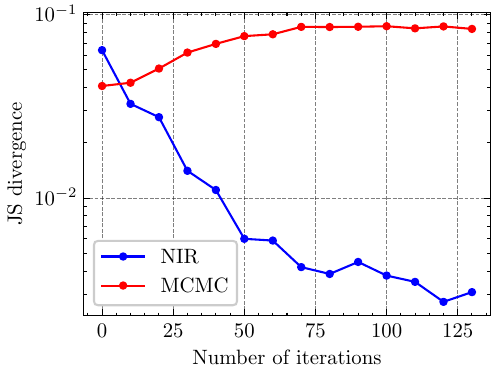}
    \caption{Evolution of the JSD between empirical configuration probabilities from the MCMC and NIR samplers and the exact probabilities during NQS optimization for a $2\times3$ lattice and $g=0.01$.} \label{fig:js_divergence}
\end{figure}

Since the only difference between the runs is the sampling method, and the divergence appears in the ferromagnetic regime, this strongly suggests that the issue arises because of MCMC sampling failure. To further support this, we compare the samplers against exact sampling on a small $3\times2$ lattice with $g=0.01$. We note that with the multi-state formulation used to target the three lowest energy states, this system effectively has 18 sites and is near the practical limit for exact sampling. We perform the comparison by sampling 25600 configurations using either the NIR or MCMC sampler, estimating the empirical configuration probabilities from these samples, and then computing the Jensen–Shannon divergence (JSD) relative to the exact probabilities obtained by evaluating the NQS over all configurations. The evolution of JSD during NQS optimization is shown in Fig. \ref{fig:js_divergence}. Early in training, the values are similar, but as the NQS approaches the lowest energy states, the JSD for MCMC increases because the distribution becomes difficult to sample locally, while the JSD for NIR decreases as the proposal network learns to approximate the NQS distribution. In this experiment, we disabled the adaptive proposal-network training described in Sec. \ref{sec:adaptive_retraining}. Instead, we used a fixed number of proposal samples and performed one proposal network update per NQS update to simplify the JSD dynamics. For such a small system, the overlap between the MCMC and true distributions remains sufficient for successful NQS training. However, as lattice size increases, the locality of the MCMC sampler becomes increasingly problematic, and the advantage of NIR becomes clearly visible in the NQS training dynamics.

\subsection{Comparing NIR with DMRG}

\begin{figure}
    \centering
    \includegraphics[width=\textwidth]{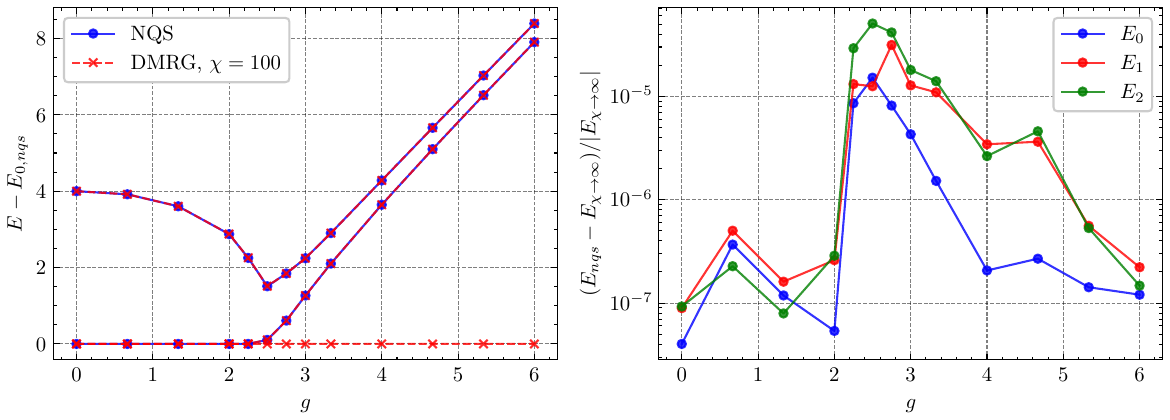}
    \caption{ \textbf{Left:} Dependence of the three lowest energy levels (relative to the ground state found with NQS) on $g$ for TFI on an $8\times 8$ lattice. Computed using NQS (blue) and DMRG (red). \textbf{Right:} Dependence of the difference between energies obtained with NQS and DMRG on $g$.} \label{fig:energies_vs_g}
\end{figure}

To demonstrate that our method performs competitively in the critical regime, we compute the dependence of the three lowest energy levels on $g$ in the range from $0$ to $6$ for $8\times8$ lattice. Fig. \ref{fig:energies_vs_g} presents the results and compares them with those obtained using the density matrix renormalization group (DMRG) \multicite{dmrg_Schollwock2011, tensor_nets_Orus2019} method, with the virtual bond dimension set to $100$. We used TeNPY \cite{tenpy} package to implement DMRG. The critical regime poses a significant general challenge for numerical studies as the correlation length diverges and the system becomes gapless, making the convergence slower. In particular, DMRG techniques rely on the low-entanglement approximation, and in large systems, they would need exponentially larger bond dimensions. On the absolute scale, both methods show good agreement and capture the closing and reopening of the energy gap for \ $g \in (2, 3)$. 

To benchmark our results more precisely, we computed energies using four DMRG bond dimensions (25, 50, 100, and 200) and extrapolated to infinite bond dimension by fitting $E(\chi) = a + b  (\chi + c)^d$, where $a$, $b$, $c$, and $d$ are fitted parameters. The right panel of Fig. \ref{fig:energies_vs_g} shows the energy error relative to these extrapolated values. The agreement is on the order of $~10^{-7}$ at low and high $g$, and on the order of $~10^{-5}$ near the critical region. 

\subsection{Benchmarking on Heisenberg $J_1$-$J_2$ model}

In this section, we benchmark the NIR sampler by searching for the ground state of the $J_1$-$J_2$ Heisenberg model at the point of maximum frustration, $J_2 = J_1 / 2$. This problem has become a standard benchmark in the NQS literature. The Hamiltonian is
\begin{equation}
\hat{H} = J_1 \sum_{\langle i,j \rangle} \vec{S}_i \cdot \vec{S}_j + J_2 \sum_{\langle\langle i,j \rangle\rangle} \vec{S}_i \cdot \vec{S}_j \,,
\end{equation}
where $\vec{S}_j$ are spin-$\frac{1}{2}$ operators defined on the sites of 2D square lattice with periodic boundary conditions. The model includes competing antiferromagnetic interactions between nearest neighbors $\langle i,j \rangle$ and next-nearest neighbors $\langle\langle i,j \rangle\rangle$, with interaction strengths $J_1$ and $J_2$, respectively.

For this experiment, we use a residual convolutional NQS architecture \cite{resnet_2016}. The input is first passed through a single convolutional layer to increase its channel dimension to 16. It is then processed by four residual blocks, each consisting of two sequential convolutional layers with an identity skip connection. Afterward, global average pooling is applied, and a final linear layer maps the result to two values representing the real and imaginary parts of $\log \psi (s)$. All convolutional layers use $3\times3$ kernels with circular padding to preserve spatial dimensions, leave the channel dimension unchanged. Each convolutional layer is followed by layer normalization \cite{layer_norm} and a GELU activation function \cite{gelu_activation}.
 
Due to circular padding and global pooling, the network is translationally invariant. We enforce spin-inversion symmetry by flipping all spins in the input if the first spin is down. Rotation and reflection symmetries of the Hamiltonian are enforced by applying the corresponding transformations to the input configurations and averaging the outputs. We also restrict the Hilbert space to the zero-magnetization sector by subtracting 30 from the real part of $\log \psi (s)$ when a configuration has non-zero magnetization. This prevents NIR from sampling those states, even though the proposal network does not explicitly enforce this constraint.

The proposal network architecture and NIR algorithm are the same as in the TFI experiments. We do not enforce symmetries or magnetization constraints in the proposal network. In principle, one could apply random symmetry-group transformations to sampled configurations to make NIR respect the symmetries, but we found this unnecessary and did not add it for simplicity.

We performed multiple runs on the $10\times10$ using the same NQS architecture but different proposal network sizes. The results are summarized in Table \ref{tab:j1j2_sampling}, which lists the proposal network embedding dimension, number of transformer layers, total number of learnable parameters, final variational energy per site, and the average sampling efficiency over the run. The runs in the first two rows were not completed because their sampling efficiency was too low, making them computationally impractical. In these cases, the proposal network failed to approximate the distribution well enough. The remaining runs give nearly identical final results, despite the proposal networks achieving different approximation accuracies. This is enabled by the adaptive scheme described in Sec. \ref{sec:adaptive_retraining}, which compensates for lower overlap between distributions by increasing the number of proposal samples.

\begin{table}[!h]
\begin{center}
\begin{tabular}{ |c|c|c|c|c| }
 \hline
 $d_\mathrm{embd}$ & $n_\mathrm{layer}$ & $n_\mathrm{param}$ & $E/N$ & $ESS / n_\mathrm{proposed}$ \\
 \hline
 16 & 2 & 7938 & - & 0.01 \\
 16 & 3 & 11154 & - & 0.03 \\
 16 & 4 & 14370 & -0.4969 & 0.10 \\
 32 & 2 & 30210 & -0.4966 & 0.15 \\
 32 & 3 & 42786 & -0.4968 & 0.21 \\
 32 & 4 & 55362 & -0.4969 & 0.28 \\
 \hline
\end{tabular}
\caption{\label{tab:j1j2_sampling} Table for final variational energies and sampling efficiencies for different sizes of proposal network.}
\end{center}
\end{table}

To our knowledge, some of the lowest energies per site reported for the $J_1$-$J_2$ model on the $10\times10$ square lattice at maximum frustration point are -0.497627 \cite{minSR_Chen_2024}, -0.497629 \cite{j1j2_rbm_pp_2021}, and -0.497634 \cite{rende2024simple} (see Table 1 in \cite{rende2024simple} for a broader comparison). Our result is slightly higher, but the goal here is not to set a new state of the art. Rather, it is to demonstrate that the NIR sampler does not break down when applied to a highly frustrated system and therefore serves as a practical alternative for general NQS applications.

\section{Conclusions}
\label{sec:conclusions}

We have introduced NIR, a sampling approach for neural quantum states that overcomes some key limitations of both MCMC and autoregressive NQS methods. By decoupling the sampling process from the NQS architecture and employing a dedicated, trainable autoregressive neural network, NIR enables efficient and robust sampling via importance resampling. This approach avoids the architectural constraints of autoregressive NQS while mitigating the mixing and equilibration issues associated with MCMC. Through numerical experiments on the transverse-field Ising and $J_1$-$J_2$ Heisenberg models, we have demonstrated that NIR achieves stable training even in regimes where MCMC fails and produces competitive results with DMRG across a wide parameter range. Our work suggests that NIR is a practical and scalable alternative for sampling in NQS-based variational algorithms, particularly in the context of multi-state NQS where traditional sampling approaches may fall short.

\section*{Acknowledgments}
This project has received funding from the Research Council of Lithuania (LMTLT), agreement No. S-ITP-24-6.

\appendix

\section{Hyperparameters}
\label{sec:appendix_hyperparams}

The following table contains a list of hyperparameters used in numerical experiments with TFI model.

\begin{table}[!h]
\begin{center}
\begin{tabular}{ |c|c|c| }
 \hline
 \textbf{Hyperparameter} & \textbf{symbol} & \textbf{value} \\
 \hline
 number of MLP hidden layers & $n_{\mathrm{layer}}$ & $4$ \\
 number of neurons in MLP hidden layers & $d_{\mathrm{mlp}}$ & $128$ \\
 proposal network embedding dimension & $d_{embd}$ & $32$ \\
 proposal network attention heads & - & $4$ \\
 proposal network number of transformer layers & - & $4$ \\
 NQS learning rate & - & $10^{-3}$ \\
 proposal network learning rate & - & $10^{-3}$ \\
 NQS training steps & - & $10,000$ \\
 Batch size (both NQS and proposal network) & - & $512$ \\
 ESS threshold & $\alpha_{ESS}$ & $2.0$ \\
 sampling efficiency threshold & $\alpha_\mathrm{eff}$ & $0.1$ \\
 MCMC warmup steps & - & 640 \\
 MCMC thinning interval & - & 10 \\
 \hline
\end{tabular}
\caption{\label{hyperparams_table} Hyperparameters table.}
\end{center}
\end{table}

\clearpage

\bibliography{references}

\end{document}